\documentclass[letter]{aa} 
\usepackage{graphicx}
\usepackage{txfonts}
\begin{document}
   \title{Radio astrometry with chromatic AGN core positions}

   \author{R. W. Porcas}

   \institute{Max-Planck-Institut f\"ur Radioastronomie,
             Auf dem H\"ugel 69, D 53121, Bonn\\
              \email {porcas@mpifr-bonn.mpg.de}
                 }
   \date{Received 8 July 2009; accepted 25 August 2009}

 
  \abstract
    {}
   {The effect of frequency-dependent AGN core positions (``core-shifts'') 
    on radio Very Long Baseline Interferometry (VLBI)
    global astrometry measurements is investigated.}
   {The basic equations relating to VLBI astrometry are reviewed, including the
   effects of source structure. A power-law representation of core-shifts, based
   on both observations and theoretical considerations of jet conditions, is incorporated.}
   {It is shown that,
in the presence of core-shifts, phase
and group-delay astrometry measurements yield different positions. For
a core displacement from the jet base parametrized by
 $\Delta${\bf x}$(\lambda) = k\lambda^{\beta}$,
group delays measure a ``reduced'' core-shift of
$(1-\beta)\Delta${\bf x}$(\lambda)$.
For the astrophysically-significant case of $\beta = 1$, group delays
measure no shift at all, giving the position of the jet base.
At 8.4\,GHz an estimated typical offset between phase
and group-delay positions of $\sim$170\,$\mu$as is smaller
than the current $\sim$250\,$\mu$as precision of group-delay positions 
of the sources used to define the ICRF; however,
this effect must be taken into account for
future measurements planned with improved accuracy
when comparing with optical positions of AGN to be obtained with the GAIA mission.}
   {}


   \maketitle
%

\section{Introduction}

In astronomy the highest angular resolution achieved on
a routine basis is provided by the technique of radio
Very Long Baseline Interferometry (VLBI).
Observations with intercontinental baselines
can be used to image sources with sub-mas
resolution, and astrometry with a precision $\sim$2 orders of
magnitude smaller is possible.
VLBI positions of a few hundred point-like extragalactic radio sources
are used to define the International
Celestial Reference Frame
(ICRF; see Ma et al. 1998; Fey et al. 2004).
Most radio sources are not points on this angular scale, however,
and are far from circularly-symmetric,
the emission rather arising from collimated jets of relativistic plasma.
Furthermore, different regions of the emission have different
spectral indices.
The question as to what points these positions refer to
is not just of academic interest.
The ESA space mission GAIA
(e.g. Lindgren et al. 2007),
due for launch in 2011, will measure the positions of many
AGN at optical wavelengths with precisions as good as 24\,$\mu$as.
Furthermore, proposed improvements in VLBI instrumentation 
(Petrachenko 2006)
hold promise for much improved VLBI global astrometric precision.
Alignment of a GAIA-based optical reference frame with the ICRF
will require a physical understanding of the location of the radio
positions at this level.

\subsection{Astrometry using interferometer phase}

The high resolution of VLBI results from the fact that
the visibility phase, $\phi$, is sensitive to changes in the geometric path
length difference in the two arms of an interferometer. For a baseline
of length $L$ this difference is $L\,cos{\theta}$, where $\theta$ is
the (instantaneous) angle between the source position, {\bf x},
and the baseline direction in the sky.
For a small position offset in the sky, $\Delta${\bf x},
the component in the resolution direction
(the direction of arc $\theta$ on the sky), $\Delta{x^{'}}$,
gives rise
to a change in path difference of
$L\,sin{\theta}\,\Delta{x^{'}}$
resulting in an ``astrometric'' phase change

\vspace{0.7mm}
$\Delta{\phi}_{astr} = (2\pi\nu/c)L\,sin{\theta}\,\Delta{x^{'}}$

\vspace{0.7mm}
\noindent
where $L\,sin{\theta}$ is the baseline length projected in the
source direction,
$\nu$ is the observing frequency and $c$ is the speed of light.
The resolution is characterised by the lobe (or fringe)-spacing,
$\Omega_{lobe} = c/({\nu}L\,sin{\theta})$,
the position change in the resolution direction which
changes $\phi$ by one turn.
For baseline lengths of several thousands of kilometers
and observing frequencies above 5\,GHz, $\Omega_{lobe}$ can be
less than 1\,mas.
Radio source images with resolutions (beam FWHM) $\sim 0.4\,\Omega_{lobe}$ are
made using VLBI arrays, and the separation between unresolved features
in such images can be determined to a small fraction of the beamwidth.
The {\it relative} position of two {\it point} sources close together on
the sky can also be directly measured using the
difference of their interferometer phases.
For measurements with high signal-to-noise ratios the precision can be
a small fraction of $\Omega_{lobe}$, $\sim$\,a few\,$\mu$as.

However, $\phi$ is only measured modulo $2\pi$
whereas the geometric path length difference, $L\,cos{\theta}$,
can be hundreds of millions of wavelengths.
Furthermore, even with an accurate knowledge of instrumental phase and clock offsets,
telescope positions and
the orientation of the baseline in space, the signal propagation through
the neutral troposphere and the ionosphere above the telescopes
adds an unknown
number of turns of phase, which varies on
a short timescale.
The additional path, $A$, through the troposphere ($\sim$2m at the zenith)
is non-dispersive,
adding an additional interferometer phase
$\phi{_{trop}}=(2\pi\nu/c)(A^B-A^A)$
where the superscripts $A,B$ refer to the two telescopes.
However,
the path through the ionosphere, $I(\nu)$, is dispersive, producing a phase advance
at each telescope, and an interferometer phase change

\vspace{0.7mm}
$\phi_{ion} = -(2\pi/c)(I_1{^B}-I_1{^A})\nu^{-1}$

\vspace{0.7mm}
\noindent
where $I_1$ is the path at unit frequency.
For close source pairs (whose atmospheric paths can be considered
the same), these problems can be solved by observing both sources simultaneously,
or by rapidly switching between them, enabling their relative
phase to be tracked as the Earth rotates, and solving for the $2\pi$
phase ambiguities (Marcaide \& Shapiro 1983).
However, for very large source separations the relative phase between two
sources cannot easily be used directly for relative astrometry.


\subsection{Group-delay astrometry}

Observations over a wide frequency band can solve the phase ambiguity problem
by furnishing another variable;
since $\phi$ varies linearly with frequency for non-dispersive
(geometric and tropospheric)
signal paths, the phase slope across the band - the group delay
$\tau = (1/2\pi)d\phi/d\nu$ -
provides an unambiguous measure of the path length difference,
$Lcos{\theta}/c$,
albeit with a reduced precision which depends on the bandwidth
spanned.
A small position offset results in an astrometric delay change

\vspace{1mm}
$\Delta\tau_{astr} = (1/2\pi)d(\Delta\phi_{astr})/d\nu = L\,sin{\theta}\,\Delta{x}^{'}/c$\,.

\vspace{1mm}
\noindent
The dispersive ionospheric contribution produces a delay  of

\vspace{1mm}
$\tau_{ion}(\nu) = (I_1{^B}-I_1{^A})\nu^{-2}/c$\,.

\vspace{1mm}
\noindent
By observing simultaneously in two widely separated frequency bands
$\nu_1, \nu_2$,
the delay difference between the bands
can be used to estimate, and remove, the ionospheric contributions to
the group delay measurement at the higher frequency $\nu_2$.

\vspace{1mm}
$\Delta\tau_{ion}(\nu_2) = (\tau_{{\nu}_2} - \tau_{{\nu}_1})/(1 - ({\nu}_1/{\nu}_2)^{-2})$\,.

\vspace{1mm}
\noindent
Group delay astrometry measurements of compact radio sources
(mostly Active Galactic Nuclei, AGN) are regularly
carried out by the International VLBI Service (IVS)
in 24-hour observing runs,
using a global network of telescopes observing at 2.3 and 8.4\,GHz,
and bandwidths spanning a few hundred MHz.
After ionospheric delay correction, the
8.4\,GHz group delays are analysed using a model for the tropospheric delay,
producing fits for Earth rotation parameters, telescope coordinates
and radio source positions.
Such group delay positions of radio sources,
and with quoted errors as small as 250\,$\mu$as,
are used to define the ICRF
to an accuracy of $\sim$30\,$\mu$as.

\subsection{Astrometry with extended sources}

The visibility phase of an {\it extended} source is conventionally
divided into 2 terms: (a) that due to the path length difference
to some reference position within the source, {\bf x$_0$}, and
(b) a source structure term,
$\phi$$_{str}$,
which can be determined by evaluating the
relevant component of the
Fourier transform of a map of the source brightness distribution
centred on {\bf x$_0$}.
Note that $\phi$$_{str}$ is generally a ``baseline-dependent'' quantity, which
allows these 2 terms to be separated by self-calibration
procedures when imaging sources with a VLBI array
(Cotton 1979; Cornwell \& Wilkinson 1981; Schwab \& Cotton 1983).
All the sources used for defining the ICRF exhibit a dominant
compact feature - the ``core'' - which provides an obvious
reference position within the structure.
The difference of visibility phase between two extended sources
can be used to determine the precise separation between their
reference points by first imaging the sources, evaluating the
structure phase terms and correcting their visibility phases
(Marcaide \& Shapiro 1983).
Charlot (1990) has emphasised that this correction may also be necessary
even when using group delays for astrometry,
in those cases where  ${\phi}_{str}$
{\it varies rapidly with resolution},
giving rise to a baseline-dependent structure delay
$(1/2\pi)d{\phi}_{str}/d\nu$
across the observing band.
In principle, source structure corrections should be made separately at
both 2.3 and 8.4\,GHz when ionospheric corrections are made to
group delays (Petrov 2007).

When sources are observed using a wide bandwidth, the change of
source structure {\it with frequency across the band} needs to be taken into account.
The technique of Multi-Frequency Synthesis (MFS) has been developed
for astronomical imaging of extended sources, which uses wide (30\%) bandwidths
in order to fill in the range of resolutions covered
by interferometer arrays
(Conway et al. 1990; Sault \& Wieringa 1994).
The algorithms used for MFS are primarily concerned with producing
a single-frequency image with high dynamic range, but they also
estimate the spectral index distribution.
Evaluating the source structure contribution to the
wideband visibility function used for group-delay measurements
in principle requires
both of these components.

\section{Frequency-dependent core-shifts}

The sources used by the IVS for defining the ICRF
are chosen to be largely point-like at both 2.3 and 8.4\,GHz,
making structural corrections apparently unimportant.
However, due to opacity in
the dominant, compact core component at the base of the
relativistic jet, the position of the peak of radio
emission is expected to be frequency-dependent (Blandford \& K\"onigl 1979;
K\"onigl 1981).
Marcaide et al. (1985) discovered a shift of 700\,$\mu$as between the positions of
the ``cores'' of the quasar 1038+528A measured at 2.3 and 8.4\,GHz
by comparing their separations from a feature in the nearby quasar 1038+528B.
Using observations at additional frequencies, they described the position change with an
{\it ad hoc} law,
$k\lambda^{\beta}$,
with $0.7 < \beta < 2.0$.
In order to
measure such ``core-shifts'' it is necessary to correctly register the images at
different frequencies.
This can be done by assuming that another feature of the
structure has a frequency-independent position,
by phase-reference imaging using another source with a
frequency-independent position, by using the technique
of frequency phase-referencing (Middelberg et al. 2005;
Rioja et al. 2005)
or by comparing the separations of multiple gravitationally-lensed images
of a single source at different frequencies (Porcas \& Patnaik 1995; Mittal et al. 2006).

Lobanov (1998) has suggested the application of such opacity effects to
the study of the conditions within the central regions of AGNs. He measured a
value $\beta = 1.04$ for the core of the quasar 3C\,345, close
to the value of 1 expected for synchrotron self-absorption in the
regime of equipartition between jet particle and magnetic field energy
densities.
Kovalev et al. (2008) have investigated the prevalence of core-shifts in a
sample of 29 sources for which a secondary (presumed achromatic) feature
in the jet could be used as a position reference. They
report a median core-shift of 440\,$\mu$as
between 2.3 and 8.4\,GHz, with a largest value of 1400\,$\mu$as.
Although the sources used in these studies necessarily had jet
components in addition to the compact, ``chromatic'' core,
there is no reason to suppose that such core-shifts are absent
{\it even in sources which show little evidence of other structure}.
Thus one must assume that the sources used to define the ICRF (at 8.4\,GHz) in
fact have frequency-dependent positions.
The measured values of core-shifts are typically smaller than
the beamwidth of VLBI arrays used to image sources
and this effect does not, therefore, easily lend
itself to correction using the MFS techniques mentioned above.

\subsection{Effect of core-shifts on astrometry measurements}

Hitherto the effects of AGN core-shifts {\it within the observing band}
on VLBI astrometric group delay measurements have not been widely considered.
Here this effect is investigated.
For simplicity the core is considered to be a point source at each frequency,
whose position along a fixed direction depends on frequency.
Following Marcaide et al. (1985) the shift, $\Delta${\bf x}$(\lambda)$, with respect to a
reference position, {\bf x}$_0$ (the jet base), is parametrized
with a power-law,
 $\Delta${\bf x}$(\lambda) = k\lambda^{\beta}$
which can be re-written as

\vspace{0.7mm}
$\Delta{x^{'}}(\nu) =\Delta{x}_{1}^{'}\nu^{-\beta}$

\noindent
where $\Delta{x}_{1}^{'}$ is the shift in the resolution direction
at unit frequency.
The astrometric phase term then becomes

\vspace{0.7mm}
$\Delta{\phi}_{astr} = (2\pi\nu/c)L\,sin{\theta}\,\Delta{x}_{1}^{'}\nu^{-\beta}$\,.

\vspace{0.7mm}
\noindent
Relative astrometry using interferometer phase measurements refers to
the frequency-shifted core position or
the average
position of the core within that band.
The astrometric contribution to the group delay, however, becomes

\vspace{0.7mm}
$\Delta\tau_{astr} = (1/2\pi)d(\Delta\phi_{astr})/d\nu = (1-\beta)L\,sin{\theta}\,\Delta{x}_{1}^{'}\nu^{-\beta}/c$\,.

\vspace{0.7mm}
\noindent
Note that the astrometric correction from a group delay measurement responds to
a ``reduced'' core-shift of $(1-\beta)\Delta{x}_{1}^{'}\nu^{-\beta}$.
In the special case of $\beta = 1$, the group delay
{\it measures no core-shift at all}
but gives a position corresponding to the jet base,
{\bf x}$_0$.
In general, phase and group-delay positions will differ at any
frequency by an amount $\beta\Delta{x}_{1}^{'}\nu^{-\beta}$.

The presence of core-shifts interacts with dual-frequency group-delay
corrections for the ionospheric path. If not taken into account,
the delay difference between measurements at two frequencies
introduced by the core-shift,

\vspace{0.7mm}
$\Delta\tau_{astr}(\nu_{2}) - \Delta\tau_{astr}(\nu_{1}) = (1-\beta)L\,sin{\theta}\,\Delta{x}_{1}^{'}(\nu_2{^{-\beta}}- \nu_1{^{
-\beta}})/c$

\vspace{0.7mm}
\noindent
will be incorporated as part of the ionospheric correction,
$\Delta\tau_{ion}(\nu_2)$,
resulting in a further (but typically smaller) shift $\Delta ${\bf x}$_{ion}$ in the measured position at $\nu_2$:

\vspace{0.7mm}
$\Delta{x}_{ion}^{'} = (1-\beta)\Delta{x}_{1}^{'}(\nu_2{^{-\beta}}-\nu_1{^{-\beta}})/(1 - ({\nu}_1/{\nu}_2)^{-2})$\,.

\vspace{0.7mm}
\noindent
Of course, for the case of $\beta=1$, there is no measured core-shift
and hence the ionospheric correction is properly applied.

\section{Discussion}

Table 1 compares values of core-shifts and group-delay measurements
for various assumed values of $\beta$ for a source having a
core-shift between 2.3 and 8.4\,GHz
of $440\,\mu$as, the median value found by Kovalev et al. (2008).
\begin{table}
\caption{Position shifts in $\mu$as assuming various values of $\beta$,
for a source with measured 
core-shift between $\nu_1$=2.3\,GHz and $\nu_2$=8.4\,GHz of
 $440\,\mu$as, the median value found by Kovalev et al. (2008)}
\label{table:1}      
\begin{center}
\begin{tabular}{rrrrrr}
\hline
$\beta$ & $\Delta${\bf x}$_{2.3}$ & $\Delta${\bf x}$_{8.4}$ & (1-$\beta$)$\Delta${\bf x}$_{8.4}$ & $\Delta ${\bf x}$_{ion}$ & $\Delta${\bf x}$_{phase-delay}$\\
\hline
0.2 & 1928 & 1488 & 1190 & 29 & 326 \\
0.4 & 1088 & 648 & 389 & 21 & 281 \\
0.6 & 814 & 374 & 150 & 14 & 239 \\
0.8 & 682 & 242 & 48 & 7 & 201 \\
1.0 & 606 & 166 & 0 & 0 & 166 \\
1.2 & 558 & 118 & -24 & -7 & 134 \\
1.4 & 526 & 86 & -34 & -14 & 106 \\
1.6 & 503 & 63 & -38 & -21 & 80 \\
1.8 & 487 & 47 & -38 & -29 & 57 \\
2.0 & 476 & 36 & -36 & -36 & 36 \\
\hline
\end{tabular}
\end{center}

Notes: columns 2 and 3 give the deduced shifts from the jet base
 {\bf x}$_0$ at 2.3 and 8.4\,GHz; column 4 the ``reduced''
group-delay shift at 8.4\,GHz; column 5 the ``ionospheric'' shift;
column 6 the total difference between positions
measured at 8.4\,GHz using interferometer phase, and ionosphere-corrected group
delays.
\end{table}
For values of $\beta$ above 0.7 the typical shift from {\bf x}$_0$ at 8.4\,GHz
is within the current quoted 250\,$\mu$as error for
group-delay positions.
For $\beta=1$ the typical shift is 166\,$\mu$as.
For a baseline length of 6000\,km (such as that between Westford, Ma, USA
and Wettzell, Germany) and an observing frequency of 8.4\,GHz, $\Omega_{lobe}$ is 1.2\,mas.
The shift produces an astrometric phase term
$\Delta{\phi}_{astr} = 49^o$.
For a {\it fixed} offset of 166\,$\mu$as the phase change
across a 720\,MHz band (as used for IVS observations)
due to the changing resolution
would amount to $4^o$. However, the small shift in
core position {\it across the band} of 14\,$\mu$as
exactly cancels this change, resulting in
a phase offset of $49^o$ {\it at all frequencies}
and an astrometric group delay $\Delta\tau_{astr} = 0$.
Note that, for $\beta > 1$, the group delay position
is on the opposite side of {\bf x}$_0$ from the jet direction.
For the (probably physically unlikely) case of $\beta = 2$
the effect of the core-shift on group delays is indistinguishable
from that of the ionosphere, and ionosphere-corrected delays
again refer to {\bf x}$_0$, as in the case of $\beta = 1$.

\subsection{The ICRF}

The ICRF is the current best realization of the Celestial Reference
Frame, based on ionosphere-corrected, 8.4\,GHz group-delay positions
for several hundred radio sources.
The sources are all compact and hopefully stable,
resulting in consistent and repeatable positions
when measured using the standard IVS observing bands.
VLBI observations link the ICRF with the International Terrestial
Reference Frame (ITRF) via group-delay positions of AGN cores in the sky
and radio telescope positions on the Earth.
Just as the mathematically well-defined, but sometimes physically
inaccessible, positions of telescopes (the intersection of the azimuth
and elevation axes)
must be located with respect to the grid of geodetic monuments on the
Earth, so too must ``mathematical'' group-delay positions of AGN cores be located
with respect to radiating points in the sky at radio and
other wavebands.
The phenomenon of core-shifts complicates this procedure although,
if $\beta = 1$ for most AGN cores, the relationship is
less frequency-dependent than one would otherwise suppose, since
all group delay positions at whatever frequency then refer to
the same point.

Jacobs \& Sovers (2008) have recently established a Celestial
Reference Frame based on VLBI group-delay measurements at
8.4 and 32\,GHz, where the 8.4\,GHz measurements are used to
make ionospheric delay corrections. Their frame is based on the
positions of 318 compact radio sources. For sources in common
with the ICRF, the weighted rms position differences are
241\,$\mu$as in RA cos(dec) and 290\,$\mu$as in declination.
These differences are close to those one might expect
from the measuring precisions alone. For the case of
$\beta = 1$, the group-delay positions at both frequencies
measure the same position of the jet base, {\bf x}$_0$.
Note that the
typical core-shift at 32\,GHz is only 44\,$\mu$as, corresponding to
the difference between phase and group-delay positions.

\subsection{Phase-reference observations}

VLBI phase-referencing is a technique whereby the visibility phase
of one source (the calibrator) is subtracted from that of another
(the target). It can be used either to remove unwanted instrumental and propagation
phase terms from a weak target source visibility, and/or to measure the relative position
between the target and calibrator.
The technique has also been successfully used for spacecraft navigation.
Calibrator sources and their positions are often chosen from those used to define the ICRF, or from
the VLBA Calibrator Survey (VCS; Petrov et al. 2008 and references therein)
for which similar
precise ionosphere-corrected 8.4\,GHz group-delay positions have been measured.
As shown above, these positions do not, in general, refer to the core position
at 8.4\,GHz. Although this does not detract at all from
the ability of phase-referencing to detect weak sources,
it does require a reinterpretation of the derived position of the target,
since this is measured with respect to the calibrator core position
at the observing frequency, not its listed group-delay position.
Knowledge of the calibrator core shift at the frequency of observation
is hence necessary to locate the target in the ICRF.

Relative positions derived from VLBI phase measurements are more
precise than positions determined using group-delays.
Fomalont (2006) has suggested that more accurate ICRF positions might be
obtained from 8.4\,GHz relative phase measurements by
first determining the relative positions of groups of sources within a radius of perhaps $20^o$,
and then ``stitching'' together many such groups to cover the whole sky.
This may indeed be possible but one should note that the relative positions
measured at 8.4\,GHz between ICRF sources will not correspond to their
relative positions measured using group-delays at 8.4\,GHz.

Marti-Vidal et al. (2008) have used wide-field phase-differenced observations
at 15.4\,GHz, and an atmospheric model, to
determine positions for 13 AGN cores in the S5 polar cap sample,
with quoted precisions ranging from 14--200\,$\mu$as.
They determined mean position corrections with respect to ICRF
group-delay positions of 278\,$\mu$as in RA and 170\,$\mu$as in declination.
Note that, for $\beta = 1$, the group-delay positions refer to the
jet base, whereas the 15.4\,GHz interferometer phase positions refer
to the 15.4\,GHz core position, typically $\sim$93\,$\mu$as away.

\subsection{VLBI2010}

Within the program VLBI2010 the IVS community is planning a major upgrade
in instrumentation for global geodetic and astrometric observing
(Petrachenko 2006).
Amongst other improvements it is planned to observe in 4 different
frequency bands between 2 and 18\,GHz, both to improve the group-delay
resolution, and also to attempt the transition from group-delay to
using the full precision of interferometer phase measurements.
Hobiger et al. (2009) have considered schemes for solving the
$2\pi$ phase ambiguities which arise in interpolating the interferometer
phase between the bands, taking into account the need to solve for
the dispersive ionospheric path and core-shifts in the different
bands.
Their simulations, however,
which use phase and group-delay measurements from the 4 bands,
assume fixed core-shifts
in each of the bands and do not take account of the shift within the bands.
In fact, for the case $\beta = 1$, the core-shift has no effect on the
group-delay measurements and adds just a constant phase offset at all
4 bands, which would make the interpolation problem in the
presence of an unknown ionospheric path easier to solve.
However, for unknown values of $\beta$ and
$\Delta${\bf x}$_{1}$
phase measurements in at least 3 frequency
bands would be needed to determine these core-shift parameters.


\section{Conclusions}

It has been shown that the phenomenon of AGN frequency core-shifts,
arising from opacity effects on the radio synchrotron emission at
the base of relativistic jets, breaks the proportional relationship
between interferometer phase and group-delay measurements in
astrometry. The effect on the latter has
been investigated using a simple model of a point-source core
whose position down the jet has a power-law dependence on wavelength.
This is no doubt an over-simplification as both the exponent $\beta$
and the jet angle may change with position along the jet (Lobanov, 1998).
Nevertheless,
the results indicate that significant position differences arise at
the sub-mas level between measurements using phase and group delays.
These differences are comparable to the
current precision of source positions derived from group delay measurements
and used to establish the ICRF.
However, they will assume a greater importance in the near future
when improved radio-astronomical instrumentation, and the GAIA
mission at optical wavelengths, produce
positions with up to an order of magnitude better precision.

The core-shift coefficient, $\Delta{x}_{1}$, and power-law exponent, $\beta$, are governed by the precise
physical conditions in the synchrotron-emitting jet. Blandford \& K\"onigl (1979)
have shown that, for synchrotron self-absorption in an equi-partition
regime, $\Delta{x}_{1}$ is proportional to (luminosity)$^{2/3}$,
whereas $\beta$ can be expected to be $\sim$1. If this is indeed common,
then group-delay measurements at any frequency refer to a fixed point
at the base of the jet,
perhaps close to the centre of optical emission.
This is more simple than
one might expect.
Note that the process of delay fringe-fitting, used for most
astronomical VLBI data analysis, determines delay (but not phase) residuals
with respect to this position.
An improved Celestial Reference Frame, based on interferometer phase
measurements of core positions, would presumably align well with the
present ICRF, since the offsets between the cores and the jet bases
in the set of defining source will be randomly distributed in angle.
The VLBI2010 project holds promise for such an advance in the coming years.

\begin{acknowledgements}
I thank Bill Petrachenko for discussions on the use of visibility
phase in future VLBI2010 observing scenarios, and Andrei Lobanov for
comments on the text.
\end{acknowledgements}

\end{document}